\documentstyle[eqsecnum,aps,epsfig,twocolumn]{revtex}

\begin{document}
\title{Successive approximations for charged particle motion}
\author{G.~H.~Hoffstaetter\thanks{Georg.Hoffstaetter@desy.de}}
\address{Deutsches Elektronen--Synchrotron (DESY), Hamburg, Germany}
\maketitle

\widetext

\begin{center}
To H. Rose on the occasion of his 65th birthday
\end{center}
\begin{abstract}
  Single particle dynamics in electron microscopes, ion or electron
  lithographic instruments, particle accelerators, and particle
  spectrographs is described by weakly nonlinear ordinary
  differential equations.  Therefore, the linear part of the
  equation of motion is usually solved and the nonlinear effects are then
  found in successive order by iteration methods.
  
  When synchrotron radiation is not important, the equation can be
  derived from a Hamiltonian or a Lagrangian.  The Hamiltonian nature
  can lead to simplified computations of particle transport through an
  optical device when a suitable computational method is used.  H.
  Rose and his school have contributed to these techniques by
  developing and intensively using the eikonal method
  \cite{rose71,plies78,rose87}.  Many ingenious microscopic
  and lithographic devices were found by Rose and his group due to the
  simple structure of this method \cite{rose81,rose90a,rose98}.
  
  The particle optical eikonal method is either derived by propagating
  the electron wave or by the principle of Maupertuis for time
  independent fields.  Maybe because of the time dependent fields
  which are often required, in the area of accelerator physics the
  eikonal method has never become popular, although Lagrange methods
  had been used sometimes already in early days \cite{sturrock55}.  In
  this area classical Hamilitonian dynamics is usually used to compute
  nonlinear particle motion.  Here I will therefore derive the eikonal
  method from a Hamiltonian quite familiar to the accelerator physics
  community.
  
  With the event of high energy polarized electron beams
  \cite{barber95a} and plans for high energy proton beams
  \cite{hoffstaetter96g}, nonlinear effects in spin motion have become
  important in high energy accelerators.  I will introduce a
  successive approximation for the nonlinear effects in the coupled
  spin and orbit motion of charged particles which resembles some of
  the simplifications resulting from the eikonal method for the pure
  orbit motion.
\end{abstract}

\narrowtext

\section{Introduction}
The well known Lagrange variational principle requires
\begin{equation}
\delta\int{\cal L}dt
=\delta\int [\vec{\tilde p}\cdot\dot{\vec q}-{\cal H}] dt=0
\end{equation}
with the Lagrangian ${\cal L}$, Hamiltonian ${\cal H}$, and
generalized momenta $\vec{\tilde p}$ and coordinates $\vec q$.

In this principle all variations of $\vec q(t)$ are allowed and
therefore the Euler--Lagrange equations of motion hold,
\begin{equation}
\frac{d}{dt}\partial_{\dot{\vec q}}{\cal L}(\vec q,\dot{\vec q},t)
=\partial_{\vec q}{\cal L}(\vec q,\dot{\vec q},t)\ .
\end{equation}
For relativistic single particle motion the Lagrangian is
\begin{equation}
{\cal L}=-mc\sqrt{c^2-\dot{\vec r}^2}+e\dot{\vec r}\cdot\vec A-e\Phi
\end{equation}
where the position $\vec r(\vec q)$ is a function of the generalized
coordinates $\vec q$.  The Jacobian matrix $\underline r$ of this
function can be written in the form $\underline r=(\partial_{\vec
  q}\vec r^T)^T$ and has the elements $r_{ij}=\partial_{q_j}r_i$.
In this efficient notation $\vec r^T$ is the transpose 

\vspace{7.5cm}

\begin{minipage}[8cm]{8cm}
\hbox{}
\end{minipage}

\vspace{7.7cm}

\noindent of the $3\times
1$ matrix $\vec r$.  The Jacobian matrix of the function $\dot{\vec
  r}(\dot{\vec q})$ is also $\underline r$ since $\dot{\vec
  r}=\sum_{i=1}^3 \dot{q_i}\partial_{q_i}\vec r = \underline r
\dot{\vec q}$.

The generalized momentum is $\vec{\tilde p}=\partial_{\dot{\vec q}}{\cal
  L}={\underline r}^T(m\gamma\dot{\vec r}+e\vec A)$ and the variational
principle can thus be written as
\begin{eqnarray}
\delta\int [\vec{\tilde p}^T\underline{r}^{-1}\dot{\vec r}-{\cal H}]dt
&=&\delta\int [(m\gamma\dot{\vec r}+e\vec A)^T\underline
  r\; \underline{r}^{-1}\dot{\vec r}-{\cal H}]dt\nonumber\\
&=&\delta\int [m\gamma v^2+e\vec A^T\dot{\vec r}-{\cal H}]dt\ .
\end{eqnarray}

If only variations $\delta_{{\cal H}=E}$ are considered which keep the
total energy ${\cal H}=E$ constant, the variational principle becomes
\begin{eqnarray}
\delta_{_{{\cal H}=E}}\int [\vec{\tilde p}\cdot\dot{\vec q}-{\cal H}] dt
&=&\delta_{_{{\cal H}=E}}\int \vec{\tilde p}\cdot d\vec q\nonumber\\
=\delta_{_{{\cal H}=E}}
\int [m\gamma v^2+e\vec A^T\dot{\vec r}]dt&=&0\ .\label{eq:var}
\end{eqnarray}
The variational principle for constant total energy is called the
principle of Maupertuis.  However, in equation (\ref{eq:var}) it does
not lead to Euler--Lagrange equations of motion, since not all
variations are allowed.

\begin{figure}
\begin{center}
\begin{minipage}{7cm}
\epsfig{file=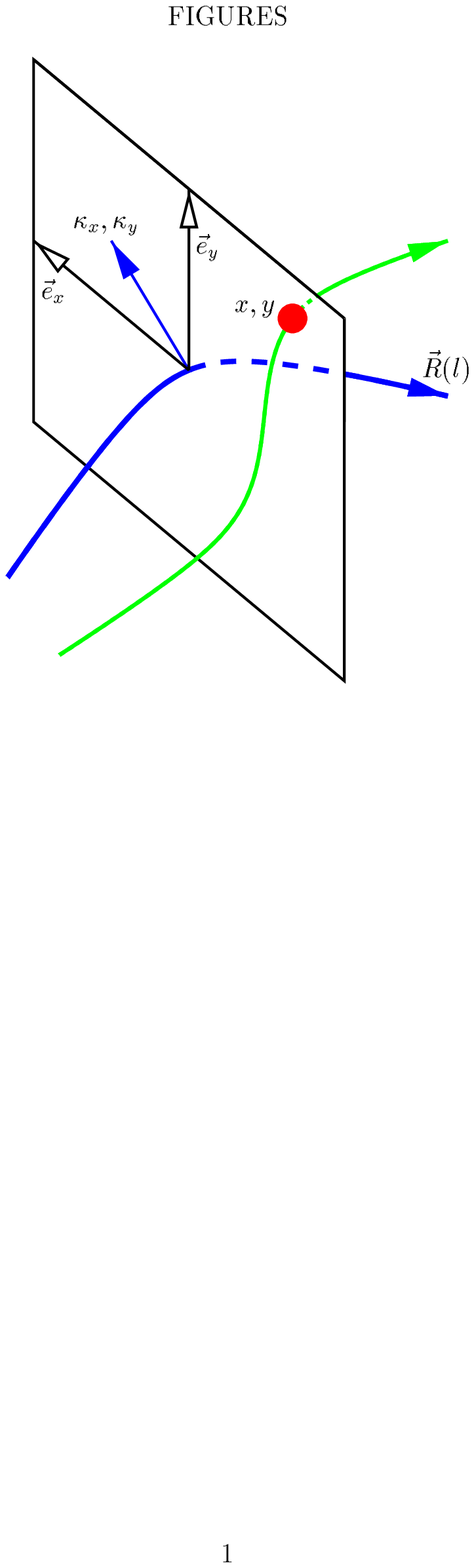,width=7cm,bb=205 470 411 747,clip}
\end{minipage}
\end{center}
\caption{Curvatures $\kappa_x$, $\kappa_y$ of the design curve and
  generalized coordinates $x$, $y$, and $l$.}
\label{fg:coord}
\end{figure}

A particle optical device usually has an optical axis or some design
curve along which a particle beam should travel.  This design curve
$\vec R(l)$ is parameterized by a variable $l$ and the position of a
particle in the vicinity of the design curve has coordinates $x$ and
$y$ along the unit vectors $\vec e_x$ and $\vec e_y$ in a plane
perpendicular this curve.  This coordinate system is shown in figure
\ref{fg:coord}.  The third coordinate vector $\vec e_l=d\vec
R/dl$ is tangential to the design curve and the curvature vector is
$\vec\kappa=-d\vec e_l/dl$.

The unit vectors $\vec e_x$ and $\vec e_y$ in the usual Frenet--Serret
comoving coordinate system rotate with the torsion of the design
curve.  If this rotation is wound back, the equations of motion do not
contain the torsion of the design curve.  The position and the
velocity are
\begin{equation}
\vec r = x\vec e_x+y\vec e_y+\vec R(l)\ ,\ \
\dot{\vec r}=\dot x\vec e_x+\dot y\vec e_y+h
\dot l\vec e_l\ ,
\end{equation}
with $h=1+x\kappa_x+y\kappa_y$.  This method is described in
\cite{rose87} and \cite{hoffstaetter95b} and is mentioned here since
design curves with torsion are becoming important when considering
particle motion in helical wigglers, undulators, and wavelength
shifters \cite{wuestefeld99}, and for polarized particle motion in
helical dipole Siberian Snakes \cite{luccio94}.

The variational principle (\ref{eq:var}) for the three generalized
coordinates $x(t)$, $y(t)$, and $l(t)$ can now be written for the two
generalized coordinates $x(l)$ and $y(l)$.  This has the following two
advantages: a) The particle trajectory along the design curve is
usually more important than the particle position at a time $t$, and
b) Whereas $\delta_{_{{\cal H}=E}}$ does not allow for all variations
of the three coordinates, the total energy can be conserved for all
variations of the two coordinates $x$ and $y$ by choosing for each
position $\vec r$ the appropriate momentum with $m\gamma
v=\sqrt{(E-e\Phi(\vec r))^2/c^2-(mc)^2}$.  We obtain from equation
(\ref{eq:var})
\begin{equation}
\delta_{_{{\cal H}=E}}\int \vec{\tilde p}\cdot d\vec q
=\delta\int [m\gamma v^2\frac{dt}{dl}+ev\vec A\cdot\frac{d\vec r}{dl}]dl=0
\end{equation}
with $d\vec r/dl=x^{\prime}\vec e_x+y^{\prime}\vec e_y+h\vec e_l$ and
$dt/dl=|d\vec r/dl|/v$.  Since all variations are allowed, the
integrand is a very simple new Lagrangian
\begin{equation}
\tilde L=m\gamma v\sqrt{{x^{\prime}}^2+{y^{\prime}}^2+h^2} 
+e(x^{\prime}A_x+y^{\prime}A_y+hA_l)
\label{eq:lag}
\end{equation}
which leads to Euler--Lagrange equations of motion
\begin{eqnarray}
{\tilde p}_x&=&\partial_{x^{\prime}}\tilde L\ ,  \ \
{{\tilde p}_x}^{\prime}=\partial_x\tilde L\ ,\\
{\tilde p}_y&=&\partial_{y^{\prime}}\tilde L\ , \ \
{{\tilde p}_y}^{\prime}=\partial_y\tilde L\ .
\end{eqnarray}
The integral $\int_0^l\tilde L(\tilde l)d\tilde l$ is called the eikonal.

Since the Hamiltonian formulation is very common in the area of
accelerator physics, we will show how the eikonal can be derived from
a Hamiltonian formulation.

The equations of motion for the three generalized coordinates $x(t)$,
$y(t)$, and $l(t)$ can be obtained from the Hamiltonian
\begin{eqnarray}
&&{\cal H}=e\Phi+\\
&&\sqrt{m^2c^2+({\tilde p}_x-eA_x)^2+({\tilde p}_y-eA_y)^2
+({\tilde p}_l/h-eA_l)^2}\ .\nonumber
\end{eqnarray}
In the case of time independent fields, ${\cal H}$ is the conserved total
energy $E$ and there are only five independent variables, rather than
six.  Note that the velocity dependent or non holonomic
\cite{goldstein80} boundary condition ${\cal H}(\vec p(\vec
q,\dot{\vec q}),\vec q,t)=E$ cannot be included in the Lagrange
formalism directly.  But in the Hamilton formalism this can be done.
Furthermore, a switch of independent variable from $t$ to $l$ can
easily be done in the Hamiltonian formulation.  The Lagrange
formulation is therefore abandoned (too easily, as will be shown
later). In the variational condition
\begin{equation}
\delta\int[\dot x{\tilde p}_x+\dot y{\tilde p}_y+\dot l
{\tilde p}_l-{\cal H}]dt=0
\end{equation}
one can change to the independent variable $l$ as follows:
\begin{equation}
\delta\int[x^{\prime}{\tilde p}_x+y^{\prime}{\tilde p}_y
+(-t^{\prime}){\cal H}-(-{\tilde p}_l)]dl=0\ .
\end{equation}
The six canonical coordinates are now $x$, ${\tilde p}_x$, $y$,
${\tilde p}_y$, $-t$, and ${\cal H}$, and the new Hamiltonian is given
by $\tilde H=-{\tilde p}_l$ which has to be expressed as a function of
the six coordinates \cite{courant58,mais96},
\begin{eqnarray}
&&\tilde H=-h{\bigg [}eA_l+\label{eq:ham}\\
&&\sqrt{({\cal H}-e\Phi)^2-(mc^2)^2-({\tilde p}_x-eA_x)^2
-({\tilde p}_y-eA_y)^2}{\bigg ]}\ .\nonumber
\end{eqnarray}
In the Hamilton formalism it is simple to take advantage of the
fact that the total energy is conserved for time independent fields;
${\cal H}^{\prime}=\partial_t \tilde H=0$ leads to ${\cal H}=E$. Then from
the six coordinates only the first four have to be considered,
leading to the Lagrangian
\begin{equation}
\tilde L=x^{\prime}{\tilde p}_x+y^{\prime}{\tilde p}_y-\tilde H\ .
\end{equation}
From $x^{\prime}=
\partial_{{\tilde p}_x}\tilde H=\frac{h}{\sqrt{\phantom{h}}}({\tilde
  p}_x-eA_x)$, $y^{\prime}=\partial_{{\tilde
    p}_y} \tilde H=\frac{h}{\sqrt{\phantom{h}}}({\tilde p}_y-eA_y)$ where
$\sqrt{\phantom{h}}$ is the square root in $\tilde H$ one obtains
\begin{eqnarray}
\sqrt{\phantom{h}}
&=&h\sqrt{\frac{(E-e\Phi)^2-(mc^2)^2}{x^{\prime 2}+y^{\prime 2}+h^2}}
\nonumber\\
&=&m\gamma v\frac{h}{\sqrt{x^{\prime 2}+y^{\prime 2}+h^2}}\ ,\\
{\tilde p}_x&=&\frac{\sqrt{\phantom{h}}}{h}x^{\prime}+eA_x\ ,\ \
{\tilde p}_y=\frac{\sqrt{\phantom{h}}}{h}y^{\prime}+eA_y\ ,
\end{eqnarray}
\begin{equation}
\tilde L=m\gamma v\sqrt{{x^{\prime}}^2+{y^{\prime}}^2+h^2} 
+e(x^{\prime}\! A_x+y^{\prime}\! A_y+h A_l)
\end{equation}
for $m\gamma v=\sqrt{(E-e\Phi)^2/c^2-(mc)^2}$.  The very simple
Lagrangian $\tilde L$ agrees with the integrand (\ref{eq:lag}) of the
eikonal.

In the following it will be shown how the Hamiltonian and the
Lagrangian equations of motion for the particle trajectory $\vec q(l)$
can be solved in an iterative way.  We write a general equation of
motion for a coordinate vector $\vec z$ in the form
\begin{equation}
{\vec z}^{\;\prime}=\vec f^1(\vec z,l)+\vec f^{\ge 2}(\vec z,l)
\label{eq:dgl}
\end{equation}
where we assume that $\vec z=0$ is a solution of the differential equation.
Furthermore, we assume $\vec z$ to be small and let $\vec f^1$ be
linear in the coordinates.  We assume that the nonlinear part of
the equation of motion can be expanded in a Taylor series $\vec f^{\ge
  2}$.  The linearized equation of motion is solved by a trajectory
$\vec z_1(l)=\underline M(l)\vec z_i$ which depends linearly on the
initial coordinates.  For the transport matrix $\underline M(l)$ we
therefore have
\begin{equation}
\underline M^{\prime}\vec z_i=\underline{f^1}\;\underline M\vec z_i
\label{eq:mdgl}
\end{equation}
for all coordinate vectors $\vec z_i$; $\underline{f^1}$ being the
Jacobian matrix of $\vec f^1$.

One can write every solution of (\ref{eq:dgl}) as $\vec
z(l)=\underline M(l)\vec \zeta(l)$, leading to the equation of
motion
\begin{equation}
\underline M^{\prime}\vec\zeta+\underline M{\vec\zeta}^{\;\prime}
=\underline{f^1}\underline M\vec\zeta+\vec
f^{\ge 2}(\vec z)\ .\label{eq:varcons}
\end{equation}
The Taylor coefficients of $\vec\zeta(\vec z_i,l)$ with respect to the
initial coordinates $\vec z_i=\vec\zeta(0)$ are called aberration
coefficients.  With equation (\ref{eq:mdgl}) one obtains
\begin{equation}
\vec z(l)=\underline M(l)\{\vec z_i
+\int_0^l{\underline M}^{-1}(\tilde l)
\vec f^{\ge 2}(\vec z(\tilde l))\}d\tilde l\ .
\label{eq:integ}
\end{equation}

Now we assume that the general solution $\vec z(\vec z_i,l)$ can be
expanded in a power series with respect to the initial coordinates.
Then symbolizing the $j$th order Taylor polynomial with $[\ldots]_j$,
we write the orders up to $j$ as $\vec z_j=[\vec z(\vec z_i,l)]_j$,
i.e.\ we use lower indices to describe the order of $\vec z_i$. The
upper index in $\vec f$ describes the order in $\vec z$, which is in
turn a nonlinear function of $\vec z_i$.  When $\vec z_{n-1}$ is
known, one can iterate the expansion up to order $n$ with equation
(\ref{eq:integ}), since
\begin{equation}
\vec z_n=\underline M(l)\{\vec z_i
+\int_0^l{\underline M}^{-1}(\tilde l)
[\vec f^{\ge 2}(\vec z_{n-1})]_n\}d\tilde l\ .
\label{eq:iter}
\end{equation}

The zeroth order of the expansion with respect to the coordinates must
vanish, which means that the trajectory $\vec q=0$ must satisfy the
equation of motion for some momentum $p(l)$.  Additionally we require
that the vector potential on the design curve is gauged to zero. This
can always be achieved.  The canonical momentum $\vec p$ then also
vanishes for the trajectory $\vec q=0$.  It then follows that the
Hamiltonian and the Lagrangian have no components linear in the
coordinates and momenta.  When computing trajectories through a
particle optical device, it is customary to normalize the momenta to
the initial design momentum $p_0=p(0)$.  The following two dimensional
generalized coordinates are therefore used:
\begin{eqnarray}
\vec q&=&{x\choose y}\ ,\ \
\vec p={{\tilde p}_x/p_0\choose {\tilde p}_y/p_0}\ ,\\
L(\vec q,{\vec q}^{\;\prime},l)&=&\tilde L/p_0\ ,\ \
\vec p              =\partial_{{\vec q}^{\;\prime}}L\ ,\ \
{\vec p}^{\;\prime}=\partial_{\vec q}  L\ ,\nonumber\\
H(\vec q,\vec p,l)  &=&\tilde H/p_0\ ,\ \
{\vec q}^{\;\prime}= \partial_{\vec p}H\ ,\ \
{\vec p}^{\;\prime}=-\partial_{\vec q}H\ .\nonumber
\end{eqnarray}
The Euler--Lagrange equations lead to the second order differential
equations $\frac{d}{dl}\partial_{{\vec q}^{\;\prime}}L=\partial_{\vec
  q}L$ for the two dimensional vector $\vec q$.

\section{Successive approximation in terms of Hamiltonians}
In the Hamilton formalism one obtains first order equations of motion
for the four dimensional vector $\vec z^T=(q_1,q_2,p_1,p_2)$.  With
the antisymmetric matrix $\underline J$ one can write the equation of
motion as
\begin{equation}
\underline J = \left(\begin{array}{rr}{\underline 0}_2&{\underline 1}_2\\
                                     -{\underline 1}_2&{\underline 0}_2
                     \end{array}\right)\ ,\ \
{\vec z}^{\;\prime}=\underline J\partial_{\vec z}H\ ,
\end{equation}
with the $2\times 2$ identity and zero matrixes ${\underline 1}_2$ and
${\underline 0}_2$.  This structure implies special symmetries for the
transport maps $\vec{\cal M}$ of particle optics.  These maps describe
how the final phase space coordinates $\vec z_f=\vec{\cal M}(\vec
z_i)$ of a particle, after flying through an optical device, are
related to the initial coordinates $\vec z_i$.  These maps are often
weakly nonlinear and can be expanded in a Taylor expansion.  The
Hamiltonian nature implies that the Jacobian $\underline{\cal
  M}=(\partial_{\vec z}\vec{\cal M}^T)^T$ of any transport map
$\vec{\cal M}(\vec z)$ is symplectic \cite{goldstein80}, meaning that
\begin{equation}
\underline{\cal M}\;\underline J\;\underline{\cal M}^T=\underline J\ .
\label{eq:symp}
\end{equation}

For the successive approximations we separate the equation of motion
into its linear and nonlinear part,
\begin{equation}
{\vec z}^{\;\prime}=\underline J\partial_{\vec z}(H^2+H^{\ge 3})\ .
\end{equation}
After we have solved for the linear transport matrix $\vec
z_1=\underline M\vec z_i$, we can iterate by equation (\ref{eq:iter})
which takes the form
\begin{equation}
\vec z_n=\underline M\{\vec z_i+\int_0^l{\underline M}^{-1}[\underline J
\partial_{\vec z} H^{\ge 3}(\vec z_{n-1})]_nd\tilde l\}
\end{equation}
With the relation $\underline{\cal M}^{-1}\;\underline J=\underline
J\;\underline{\cal M}^T$ from equation (\ref{eq:symp}) this can be
written as
\begin{equation}
\vec z_n=\underline M\{\vec z_i+\underline J\int_0^l{\underline M}^T[
\partial_{\vec z} H^{\ge 3}(\vec z_{n-1})]_nd\tilde l\}\ .
\end{equation}
The corresponding equation for the aberrations
${\vec\zeta}_n={\underline M}^{-1}{\vec z}_n$ becomes
\begin{equation}
\vec \zeta_n=\vec z_i+\underline J\int_0^l[
\partial_{\vec\zeta} H^{\ge 3}(\underline M\vec\zeta_{n-1})]_nd\tilde l\ .
\label{eq:zeta}
\end{equation}
This form of the iteration equation is quite simple. However, since
the Hamiltonian (\ref{eq:ham}) is a complicated function, the
evaluation of the four integrals can become very cumbersome.

\section{Successive approximation in terms of Lagrangians}
In \cite{rose87} Rose used a variational principle to derive a
successive approximation to nonlinear motion based on the eikonal.
This method iterates position $\vec q$ and momentum $\vec p$ in their
nonlinear dependence on the initial position ${\vec q}_i$ and momentum
$\vec p_i$.  Knowing the order $n-1$ dependence ${\vec q}_{n-1}$ and
$\vec p_{n-1}$, one has to compute ${\vec q}^{\;\prime}_{n-1}$ by
differentiation of ${\vec q}_{n-1}$ or by inversion of $\vec
p=\partial_{{\vec q}^{\;\prime}}L(\vec q,{\vec q}^{\;\prime},l)$.
Then the eikonal can be evaluated to compute the order $n$ dependence
${\vec q}_n$ and $\vec p_n$.  In general it can be cumbersome to
compute ${\vec q}^{\;\prime}_{n-1}$ and therefore here we derive a new
version of the eikonal method, which iterates directly ${\vec q}^{\;
  \prime}_n$ rather than the momentum.

In deriving the simple form of equation (\ref{eq:zeta}), advantage has
only been taken of the symplectic first order transfer matrix.  We
therefore wish to exploit this advantage again by working with new
coordinates which are identical with the canonical $\vec q$ and $\vec
p$ up to first order so that the new coordinates lead to the same
first order transport matrix $\underline M$.  To first order one
obtains
\begin{equation}
\vec p =\partial_{{\vec q}^{\;\prime}}L
= \frac{p(s)}{p_0}{\vec q}^{\;\prime}+
\frac{e}{p_0}{A_x^1\choose A_y^1} + {\cal O}^2(\vec q,{\vec q}^{\;\prime})
\end{equation}
where $p(s)$ is the momentum of a particle traveling on the design
curve $\vec q=0$, and the upper index 1 specifies the part of the
vector potential linear in $x$ and $y$.  We therefore work with the
coordinates
\begin{equation}
\vec Q={\vec q\choose\vec u}={{\vec q}\choose{\frac{p(s)}{p_0}{\vec
  q}^{\;\prime} +\frac{e}{p_0}{{A_x^1}\choose{A_y^1}}}}\ .
\label{eq:newco}
\end{equation}
Moreover, it can be shown
\cite{plies71,rose87,hoffstaetter91a,hoffstaetter93d} that the
contribution from the vector potential can be gauged to vanish
whenever there is no longitudinal magnetic field $B_0 {\vec e}_l$ on
the design curve.  Then if one investigates trajectories which start
with momentum $p_0$ in a region free of such a field, we have the
simple relation $\vec u_i={\vec q}^{\;\prime}_i$.

By splitting the Lagrangian into its second order and its higher order
part, the equation of motion becomes
\begin{eqnarray}
{\vec Q}^{\;\prime}&=&
{\vec q\choose \vec u}^{\prime}\label{eq:lagode}\\
&=&
{\frac{p_0}{p(s)}\vec u-\frac{e}{p(s)}{A_x^1\choose A_y^1}\choose 
                      \partial_{\vec q}L^2}-{0\choose 
\frac{d}{d\tilde l}\partial_{{\vec q}^{\;\prime}}L^{\ge 3}
-\partial_{\vec q}L^{\ge 3}
}\ .\nonumber
\end{eqnarray}
After having solved the linearized equation of motion, we obtain with
equation (\ref{eq:iter})
\begin{eqnarray}
\vec Q
&=&\underline M\{{\vec Q}_i
-\int_0^l\underline M^{-1}
{0\choose 
\frac{d}{d\tilde l}\partial_{{\vec q}^{\;\prime}}L^{\ge 3}
-\partial_{\vec q}L^{\ge 3}}
d\tilde l\}\\
&=&\underline M\{{\vec Q}_i
+\underline J\int_0^l\underline M^T
{
\frac{d}{d\tilde l}\partial_{{\vec q}^{\;\prime}}L^{\ge 3}
-\partial_{\vec q}L^{\ge 3}
\choose 0
}d\tilde l\}\ .\nonumber
\end{eqnarray}
An integration by parts leads to
\begin{eqnarray}
\vec Q
&=&\underline M\{
{\vec Q}_i
-\underline J\int_0^l[{\underline M^{\prime}}^T
{ \partial_{{\vec q}^{\;\prime}}L^{\ge 3}\choose 0 }
+{\underline M}^T
{ \partial_{\vec q}L^{\ge 3}\choose 0 }]
d\tilde l\nonumber\\
&&+\underline J{\bigg [}
{\underline M}^T{\partial_{{\vec q}^{\;\prime}}L^{\ge 3}\choose 0}
{\bigg ]}_0^l\}\ .\label{eq:lag0}
\end{eqnarray}
Writing the Jacobian as ${\underline M}^T=\partial_{{\vec Q}_i}{\vec
  Q}_1^T=\partial_{{\vec Q}_i}(\vec Q-{\vec Q}_{\ge 2})^T$ where $\vec
Q={\vec Q}_1+{\vec Q}_{\ge 2}$ was split into parts which depend on ${\vec
Q}_i$ linearly and nonlinearly, we obtain
\begin{eqnarray}
&&{\underline M}^{-1}\vec Q
={\vec Q}_i
-\underline J\int_0^l[
(\partial_{{\vec Q}_i}({\vec q}^{\;\prime T}-{\vec q}_{\ge 2}^{\;\prime T}))
 \partial_{{\vec q}^{\;\prime}}L^{\ge 3}\nonumber\\
&&+(\partial_{{\vec Q}_i}({\vec q}^T-{\vec q}_{\ge 2}^T))
 \partial_{\vec q}L^{\ge 3}]d\tilde l+\underline J{\bigg [}
(\partial_{{\vec Q}_i}{\vec q}_1^T)\partial_{{\vec q}^{\;\prime}}L^{\ge 3}
{\bigg ]}_0^l\nonumber\\
&&=
{\vec Q}_i
-\underline J\int_0^l[\partial_{{\vec Q}_i}L^{\ge 3}
-(\partial_{{\vec Q}_i}{\vec q}_{\ge 2}^{\;\prime T})
\partial_{{\vec q}^{\;\prime}}(L-L^2)\nonumber\\
&&-(\partial_{{\vec Q}_i}{\vec q}_{\ge 2}^T) \partial_{\vec q}(L-L^2)]d\tilde l
+\underline J{\bigg [}
(\partial_{{\vec Q}_i}{\vec q}_1^T)\partial_{{\vec q}^{\;\prime}}L^{\ge 3}
{\bigg ]}_0^l\nonumber\\
&&=
{\vec Q}_i
-\underline J\int_0^l[\partial_{{\vec Q}_i}L^{\ge 3}
+(\partial_{{\vec Q}_i}{\vec q}_{\ge 2}^{\;\prime T})
\partial_{{\vec q}^{\;\prime}}L^2
+(\partial_{{\vec Q}_i}{\vec q}_{\ge 2}^T) \partial_{\vec q}L^2
\nonumber\\
&&-(\partial_{{\vec Q}_i}{\vec q}_{\ge 2}^{\;\prime T})
\partial_{{\vec q}^{\;\prime}}L
-(\partial_{{\vec Q}_i}{\vec q}_{\ge 2}^T)\frac{d}{d\tilde l}
\partial_{{\vec q}^{\;\prime}}L]d\tilde l
\nonumber\\
&&+\underline J{\bigg [}
(\partial_{{\vec Q}_i}{\vec q}_1^T)\partial_{{\vec q}^{\;\prime}}L^{\ge 3}
{\bigg ]}_0^l\ .
\label{eq:lag1}
\end{eqnarray}
Note that the $A'_x$, $A'_y$, $p(s)$ and $p_0$ of equation
(\ref{eq:newco}) drop out of the right hand side of equation
(\ref{eq:lag1}) owing to the multiplications by the zeros in equation
(\ref{eq:lag0}).  The second order $L^2$ of the Lagrangian is a
quadratic form in which every quadratic combination of the $\vec q$
and ${\vec q}^{\;\prime}$ can occur.  It can be written using a matrix
$\underline{L^2}$ as $L^2={\vec Q}^T\underline{L^2}\vec Q$. Part of
the above integrand can be rewritten as
\begin{equation}
 {\vec q}_{\ge 2}^{\;\prime T}\partial_{{\vec q}^{\;\prime}}L^2
+{\vec q}_{\ge 2}^T \partial_{\vec q}L^2
={\vec Q}_{\ge 2}^T\underline{L^2}\vec Q
+{\vec Q}^T\underline{L^2}{\vec Q}_{\ge 2}\ .
\end{equation}
For convenience we write $L^2(\vec a)={\vec a}^T\underline{L^2}\vec
a$.  Another integration by parts in equation (\ref{eq:lag1}) leads to
\begin{eqnarray}
{\underline M}^{-1}\vec Q&=&
{\vec Q}_i
-\underline J\int_0^l[\partial_{{\vec Q}_i}(
L^{\ge 3}+L^2({\vec Q}_{\ge 2}))
\nonumber\\
&&+(\partial_{{\vec Q}_i}{\vec q}_{\ge 2}^{\;\prime T})
\partial_{{\vec q}^{\;\prime}_1}L^2({\vec Q}_1)
+(\partial_{{\vec Q}_i}{\vec q}_{\ge 2}^{T})
\partial_{{\vec q}_1}L^2({\vec Q}_1)]
d\tilde l\nonumber\\
&&+\underline J{\bigg [}
(\partial_{{\vec Q}_i}{\vec q}_{\ge 2}^T)\partial_{{\vec q}^{\;\prime}}L
+(\partial_{{\vec Q}_i}{\vec q}_1^T)\partial_{{\vec q}^{\;\prime}}L^{\ge 3}
{\bigg ]}_0^l\ .
\end{eqnarray}
The first part of the integral contains $L_E=L^{\ge 3}+L^2({\vec
  Q}_{\ge 2})$.  The integral $\int_0^l L_Ed\tilde l$ is called the
perturbation eikonal.  This scheme embodies the essential requirement
that the $n+1$ order dependence of $L_E$ on the initial variables
$\vec Q_i$ can be computed already when $Q_{n-1}$ is known; $\vec Q_n$
does not need to be known.  For an iteration of ${\vec Q}_n$,
knowledge of ${\vec Q}_{n-1}$ is sufficient.  Since ${\vec q}_1$
satisfies the first order equation of motion, we can use the relation
$\partial_{{\vec q}_1}L^2({\vec Q}_1) = \frac{d}{dl}\partial_{{\vec
    q}^{\;\prime}_1}L^2({\vec Q}_1)$ to perform another integration by
parts,
\begin{eqnarray}
{\underline M}^{-1}\vec Q&=&
{\vec Q}_i
-\underline J\partial_{{\vec Q}_i}\int_0^lL_E d\tilde l
+\underline J{\bigg [}
(\partial_{{\vec Q}_i}{\vec q}_{\ge 2}^T)\partial_{{\vec q}^{\;\prime}}L
\nonumber\\
&&+
(\partial_{{\vec Q}_i}{\vec q}_1^T)\partial_{{\vec q}^{\;\prime}}L^{\ge 3}
-(\partial_{{\vec Q}_i}{\vec q}_{\ge 2}^{T})
\partial_{{\vec q}^{\;\prime}_1}L^2({\vec Q}_1)
{\bigg ]}_0^l\nonumber\\
&=&
{\vec Q}_i
-\underline J\partial_{{\vec Q}_i}\int_0^lL_E d\tilde l
+\underline J{\bigg [}
(\partial_{{\vec Q}_i}{\vec q}^T)\partial_{{\vec q}^{\;\prime}}L
\nonumber\\
&&-
(\partial_{{\vec Q}_i}{\vec q}_1^T)\partial_{{\vec q}^{\;\prime}}L^2
-(\partial_{{\vec Q}_i}{\vec q}_{\ge 2}^{T})
\partial_{{\vec q}^{\;\prime}_1}L^2({\vec Q}_1)
{\bigg ]}_0^l\ .
\end{eqnarray}
$\partial_{{\vec Q}_i}L_E$ is the part of $\partial_{{\vec Q}_i}L$
which up to order $n$ in ${\vec Q}_i$ does not depend on ${\vec Q}_n$.
Similarly the term outside the integral is simply the part of
$(\partial_{{\vec Q}_i}{\vec q}^T)\partial_{{\vec q}^{\;\prime}}L$
which up to order $n$ does not depend on ${\vec Q}_n$.  We therefore
write $(\partial_{{\vec Q}_i}{\vec q}^T)\partial_{{\vec q}^{\;
    \prime}}L- (\partial_{{\vec Q}_i}{\vec q}_1^T)\partial_{{\vec
    q}^{\;\prime}}L^2 -(\partial_{{\vec Q}_i}{\vec q}_{\ge 2}^{T})
\partial_{{\vec q}^{\;\prime}_1}L^2({\vec Q}_1)=\{(\partial_{{\vec
    Q}_i}{\vec q}^T)\partial_{{\vec q}^{\;\prime}}L\}_E$.  If we now
express the Lagrangian in terms of the aberrations $\vec\xi$ with
$\vec Q=\underline M\vec\xi$, we obtain the iteration equation
\begin{eqnarray}
{\vec\xi}_n=_n
{\vec Q}_i
&-&\underline J\partial_{{\vec Q}_i}\int_0^l
L_E(\underline M\vec\xi_{n-1})d\tilde l
\label{eq:eikiter}\\
&+&\underline J
{\bigg [}
\{(\partial_{{\vec Q}_i}{\vec q}^T)\partial_{{\vec q}^{\;\prime}}L\}_E
{\bigg ]}_0^l\ .\nonumber
\end{eqnarray}
When computing ${\vec\xi}_n$ from ${\vec\xi}_{n-1}$ with this
iteration equation, all parts of the right hand side which contribute
to higher orders are neglected, as indicated by $=_n$.  This iteration
equation can have several advantages over the Hamiltonian iteration
equation (\ref{eq:iter}):
\begin{itemize}
\item[a)] The Lagrangian (\ref{eq:lag}) is a much simpler function
  than the Hamiltonian (\ref{eq:ham}).
\item[b)] The derivative in equation (\ref{eq:eikiter}) is performed
  after the integral has been evaluated.  Therefore only one integral
  has to be computed and it describes all four coordinates of
  ${\vec\xi}_n$.
\item[c)] The fact that the various coordinates are the derivatives
  with respect to initial conditions yields very simple relations
  \cite{hoffstaetter98f} between the various expansion coefficients of
  ${\vec\xi}_n$, which are the so called aberration coefficients of
  particle optical devices.  These relations can be much simpler than
  relations entailed by the symplectic symmetry implicit in the
  Hamiltonian formulation.
\item[d)] The second pair of coordinates in equation (\ref{eq:newco})
  can be calculated very easily.  With equations (\ref{eq:varcons})
  and (\ref{eq:lagode}), the equation of motion for $\vec \xi$ is
\begin{equation}
  \underline M{\vec\xi}^{\;\prime}=
  {0\choose 
  \frac{d}{d\tilde l}\partial_{{\vec q}^{\;\prime}}L^{\ge 3}
-\partial_{\vec q}L^{\ge 3}
  }\ .\label{eq:zeroder}
\end{equation}
After having computed ${\vec q}_n=\vec M_{2\times 4}{\vec\xi}_n$ by
iteration, the derivative ${\vec q}^{\;\prime}$ can then easily be
computed as ${\vec q}^{\;\prime}_n={\vec M}^{\;\prime}_{2\times
  4}{\vec\xi}_n$ using equation (\ref{eq:zeroder}).  One thus only
needs to iterate the two dimensional vector ${\vec q}_n$ and not a
four dimensional vector ${\vec z}_n$ as in the Hamiltonian iteration
procedure.
\end{itemize}

\section{Successive approximation for spin orbit motion}
The time variation of a spin $\vec s$ in the rest frame of a particle
is described by the so called Thomas--BMT equation $\dot{\vec
  s}=\vec\Omega_{BMT}\times\vec s$ \cite{thomas27,bargmann59} where
\begin{eqnarray}
\vec\Omega_{BMT}=
&-&\frac{q}{m\gamma}\{(a\gamma+1)\vec B_\perp+(1+a)\vec B_{\|}\\
&-&\frac{\gamma}{c}\vec\beta\times\vec E(a+\frac{1}{1+\gamma})\}
\nonumber
\end{eqnarray}
with the electric field $\vec E$, the parts of the magnetic field
$\vec B$ which are perpendicular ($\perp$) and parallel ($\|$) to the
particle's velocity, and the anomalous gyro-magnetic factor
$a=\frac{g-2}{2}$.

Changing to the comoving coordinate system of figure (\ref{fg:coord}),
we obtain $\vec s=S_x\vec e_x+S_y\vec e_y+S_l\vec e_l$ and ${\vec
  s}^{\;\prime}=(S^{\prime}_x-S_l\kappa_x) \vec
e_x+(S^{\prime}_y-S_l\kappa_y)\vec
e_y+(S^{\prime}_l+S_x\kappa_x+S_y\kappa_y)\vec e_l$.  The equation of
motion for the vector $\vec S$ of these spin components is then
\begin{eqnarray}
{\vec S}^{\;\prime}&=&\vec\Omega\times\vec S\ ,\\
\vec\Omega
&=&\vec\Omega_{BMT}\frac{h}{v}\sqrt{x^{\prime 2}+y^{\prime 2}+h^2}
-\vec\kappa\times\vec e_l\ .\nonumber
\end{eqnarray}
The equations of motion for the phase space vector $\vec z$ and the
spin $\vec S$ have the form
\begin{equation}
{\vec z}^{\;\prime}=\vec f(\vec z,l)\ ,\ \
{\vec S}^{\;\prime} = \vec\Omega(\vec z,l)\times\vec S\ .
\end{equation}
The general solutions transporting the coordinates along the optical
system, starting at the initial values $\vec z_i$, $\vec S_i$, is
given by the transport map ${\cal M}$ and the rotation matrix
$\underline{R}\in SO(3)$,
\begin{equation}
\vec z(l)=\vec{\cal M}(\vec z_i,l)\ ,\ \
\vec S(l)=\underline{R}(\vec z_i,l)\vec S_i\ .
\end{equation}
In order to find the general solution, one could compute
the nine coefficients of the rotation matrix by solving the
differential equation
\begin{equation}
R_{ij}(\vec z_i,l)^{\prime}=\epsilon_{ilk}\Omega_l
R_{kj}(\vec z_i,l)\ ,
\end{equation}
where the vector product was expressed by the totally antisymmetric
tensor $\epsilon_{ilk}$. However, computing the nine components of the
rotation matrix seems inefficient, since a rotation can be represented
by three angles.  It has turned out \cite{hoffstaetter96d} to be most
efficient to represent the rotation of spins by the quaternion $A$
which gives the rotation transformation in the SU(2) representation as
\begin{equation}
A=a_0\underline 1-{\rm i}\vec a\cdot\vec{\underline\sigma}\ .
\label{eq:su2}
\end{equation}
Here $\underline 1$ is the $2\times 2$ identity matrix and the
elements of the vector $\vec{\underline\sigma}$ are the three two
dimensional Pauli matrixes.  When a rotation by an angle $\phi$ is
performed around the unit vector $\vec e$, the quaternion
representation of the rotation has $a_0=\cos(\phi/2)$ and $\vec
a=\sin(\phi/2)\vec e$.  Therefore $a_0^2+{\vec a}^2=1$ and the
identity transformation is represented by $a_0=1$.

If a particle traverses an optical element which rotates the
spin according to the quaternion $A$ and then passes through an
element which rotates the spin according to the quaternion $B$,
the total rotation of the spin is given by
\begin{eqnarray}
C&=&c_0\underline 1-{\rm i}\vec c\cdot\vec{\underline\sigma}
=(b_0\underline 1-{\rm i}\vec b\cdot\vec{\underline\sigma})
(a_0\underline 1-{\rm i}\vec a\cdot\vec{\underline\sigma})\nonumber\\
&=&
(b_0 a_0-\vec b\cdot\vec a)\underline 1
-{\rm i}(b_0\vec a+\vec ba_0+\vec b\times\vec a)\cdot\vec{\underline\sigma}\ .
\end{eqnarray}
The concatenation of quaternions can be written in matrix form as
\begin{eqnarray}
\vec C&=&{c_0\choose\vec c}=\underline B{a_0\choose\vec a}\ ,
\label{eq:concat}\\
\underline B&=&\left(\begin{array}{rrrr}
 b_0&-b_1&-b_2&-b_3\\
 b_1& b_0&-b_3& b_2\\
 b_2& b_3& b_0&-b_1\\
 b_3&-b_2& b_1& b_0
\end{array}\right)\ .\nonumber
\end{eqnarray}
This concatenation of two quaternions can be used to
find a differential equation for the spin rotation.

While propagating along the design curve by a distance $dl$, spins are
rotated by an angle $\Omega dl=|\Omega| dl$ around the vector
$\vec\Omega$.  After having been propagated to $l$ by the quaternion
$A$, a spin gets propagated from $l$ to $l+dl$ by the quaternion
with $b_0=1$ and $\vec b=\frac{1}{2}\vec\Omega dl$.  The resulting
total rotation is given by $A+A^{\prime}dl$ and we obtain the
differential equation
\begin{equation}
{a_0^{\prime}\choose\vec a^{\;\prime}}=
\frac{1}{2}\left(\begin{array}{rrrr}
        0&-\Omega_1&-\Omega_2&-\Omega_3\\
 \Omega_1&        0&-\Omega_3& \Omega_2\\
 \Omega_2& \Omega_3&        0&-\Omega_1\\
 \Omega_3&-\Omega_2& \Omega_1&        0
\end{array}\right)
{a_0\choose\vec a}\ .
\end{equation}
Writing the matrix as $\underline\Omega$ and the vector as $\vec A$,
the spin orbit equation of motion has the form
\begin{equation}
{\vec z}^{\;\prime}=\vec f(\vec z,l)\ ,\ \ \vec A^{\prime}
=\underline\Omega(\vec z,l)\vec A\ .
\label{eq:quatdgl}
\end{equation}
The starting conditions are $\vec z(0)=\vec z_0$, $a_0=1$, and
$\vec a=0$.  The quaternion $A$ depends on the initial phase space
coordinates $\vec z_i$ and can be expanded in a Taylor series with
respect to these coordinates.  In the following we want to devise an
iteration method for $A_n$, which is the Taylor expansion to
order $n$ of $A$.

The rotation vector $\vec\Omega$ is split into its value on the design
curve and its phase space dependent part as $\vec\Omega(\vec z,l)=
\vec\Omega^0(l)+\vec\Omega^{\ge 1}(\vec z,l)$.  The spin motion on the
design curve is given by $\vec A^{\prime}_0(l)=\underline\Omega^0\vec
A_0(l)$. Similarly to equation (\ref{eq:varcons}), spin aberrations
are defined with respect to the leading order motion. Small phase
space coordinates will create a rotation which differs little from
$\vec A_0(l)$ and we write the phase space dependent rotation as a
concatenation of $\vec A_0$ and the $\vec z$ dependent rotation
$(1+\delta,\vec\delta)$ which reduces to the identity for $\vec z=0$
by requiring that the aberrations $\delta$ and $\vec\delta$ vanish on
the design curve.  With equation (\ref{eq:concat}) we obtain
\begin{equation}
\vec A=\underline{A_0}{1+\delta\choose\vec\delta}\ .
\end{equation}
The quaternion $A$ is now inserted in the differential equation
(\ref{eq:quatdgl}) to obtain
\begin{eqnarray}
\underline A^{\prime}_0{ 1+\delta\choose\vec\delta}
&+&\underline A_0{\delta^{\prime}\choose{\vec\delta}^{\prime}}\nonumber\\
&=&(\underline{\Omega}^0+\underline{\Omega}^{\ge 1})\underline A_0
{1+\delta\choose\vec\delta}\ .
\end{eqnarray}
Taking into account the equation on the design curve and the fact that
$\underline{A}_0^T$ describes the inverse rotation of
$\underline{A}_0$, we obtain
\begin{equation}
{\delta^{\prime}\choose{\vec\delta}^{\prime}}
=(\underline{A}_0^T\underline{\Omega}^{\ge 1}\underline{A}_0)
{1+\delta\choose{\vec\delta}}
=\underline{\tilde\Omega}(\vec z,l)
{1+\delta\choose{\vec\delta}}\ .
\end{equation}
Writing the Taylor expansion to order $n$ in $\vec z_i$ one finally
obtains the iteration equation
\begin{equation}
{\delta_n\choose{\vec\delta}_n}
=_n\int_0^l
\underline{\tilde\Omega}(\vec z_{n})
{1+\delta_{n-1}\choose{\vec\delta}_{n-1}}d\tilde l\ ,\ \
{\delta_0\choose\vec\delta_0}=0\ .
\end{equation}
This iteration method was used for the spin transport in the program
SPRINT \cite{hoffstaetter96d} and was evaluated using MATHEMATICA in
\cite{weissbaecker98}.

In the case of successive approximation in terms of the Hamiltonian,
the various aberration coefficients were related by the symplectic
symmetry.  With the Lagrange formalism the various aberration
coefficients were related by their being derivatives of a common
integral with respect to different initial coordinates.  In the case
of the successive approximation for spin motion, the various
aberration coefficients in $\vec\delta$ and $\delta$ are related by
the relation $(1+\delta)^2+\vec\delta^2=0$.

\subsection*{Acknowledgment}
I owe thanks to D.~Barber, H.~Mais, and M.~Vogt for thoroughly
reading the manuscript and for the resulting improvements.

\end{document}